\documentclass{natureprintstyle}

\spacing{1}
\spacing{2}
\spacing{1}

\usepackage{graphicx}
\usepackage{color}
\usepackage{bm}
\usepackage{longtable}
\usepackage{amssymb}
\usepackage{rotating}
\usepackage{hyperref}

\newcommand{\muhz}{$\mu$Hz}
\newcommand{\numax}{$\nu_{\mathrm{max}}$}
\newcommand{\dnu}{$\Delta\nu$}

\newcommand{\kepler}{\textit{Kepler}}

\newcommand{\msun}{M$_\odot$}
\newcommand{\msol}{M$_\odot$}

\newcommand{\nat}{Nature}

\let\citep\cite
\let\citet\cite

\title{A prevalence of dynamo-generated magnetic fields in the cores of
intermediate-mass stars}

\begin{document}

\author{
Dennis~Stello$^{1,2}$, 
Matteo~Cantiello$^{3}$,
Jim~Fuller$^{4,3}$, 
Daniel~Huber$^{1,5,2}$, 
Rafael~A.~Garc\'\i a$^{6}$, 
Tim~R.~Bedding$^{1,2}$, 
Lars~Bildsten$^{3,7}$, 
Victor~Silva~Aguirre$^{2}$ 
}
\maketitle

\let\thefootnote\relax\footnote{
\begin{affiliations}
\small
\item Sydney Institute for Astronomy (SIfA), School of Physics, University of Sydney, NSW 2006, Australia

\item Stellar Astrophysics Centre, Department of Physics and Astronomy,
  Aarhus University, Ny Munkegade 120, DK-8000 Aarhus C, Denmark

\item Kavli Institute for Theoretical Physics, University of California,
  Santa Barbara, CA 93106

\item TAPIR, Walter Burke Institute for Theoretical Physics, Mailcode 350-17
California Institute of Technology, Pasadena, CA 91125

\item SETI Institute, 189 Bernardo Avenue, Mountain View, CA 94043, USA

\item Laboratoire AIM, CEA/DSM -- CNRS -- Univ. Paris Diderot -- IRFU/SAp
Centre de Saclay, 91191 Gif-sur-Yvette Cedex, France

\item Department of Physics, University of California, Santa Barbara, CA 93106

\end{affiliations}
}

\begin{abstract}
Magnetic fields play a role in almost all stages of stellar evolution
\citep{Landstreet_1992}. Most low-mass stars, including the Sun, show
surface fields that are generated by dynamo processes in their
convective envelopes \citep{Parker_1955,Donati_2009}. Intermediate-mass
stars do not have deep convective envelopes \citep{Kippenhahn_1990}, although
10\% exhibit strong surface fields that are presumed to be residuals
from the stellar formation process \citep{2008CoSka..38..443P}. These
stars do have convective cores that might produce internal magnetic
fields \citep{Brun_2005}, and these might even survive into later stages
of stellar evolution, but information has been limited by our inability to
measure the fields below the stellar surface \citep{Auri_re_2015}. Here
we use asteroseismology to study the occurrence of strong magnetic
fields in the cores of low- and intermediate-mass stars. We have
measured the strength of dipolar oscillation modes, which can be
suppressed by a strong magnetic field in the core \citep{Fuller15},
in over 3600 red giant stars observed by \kepler. About 20\% of our sample show
mode suppression but this fraction is a strong function of mass. Strong
core fields only occur in red giants above 1.1 solar masses (1.1\msol), and
the occurrence rate is at least 60\% for intermediate-mass stars
(1.6--2.0\msol), indicating that powerful dynamos were very common in the
convective cores of these stars.

\end{abstract}

Red giants are formed when a low- or intermediate-mass star has finished
burning the hydrogen in its core. This leaves an inert helium core
surrounded by a thin hydrogen-burning shell and a very thick outer
convective envelope.
Like the Sun, red giants oscillate in a broad comb-like frequency spectrum of radial and non-radial
acoustic modes that are excited by the turbulent surface convection \citep{De_Ridder_2009}.
The observed power spectrum has a roughly Gaussian envelope whose central frequency, \numax, decreases as a star expands during the
red giant phase \citep{Brown_1991,Stello_2009}.
The comb structure of the spectrum arises from a series of 
overtone modes separated by the so-called large frequency separation, \dnu. One of these overtone sequences is seen for each spherical degree, $\ell$.  For observations of unresolved distant stars,
geometric cancellation prevents detection of modes with $\ell>3$.  
Their spectra are characterised by a pattern of radial ($\ell=0$) and
quadrupolar ($\ell=2$) modes that form close pairs, interspersed with dipolar
($\ell=1$) modes located roughly halfway between successive radial-quadrupolar pairs.
The octupolar modes ($\ell=3$) are weak or undetectable.
The dipolar modes have turned out to be particularly useful probes of
internal structure \citep{GarciaStello15}. 
They have been used to distinguish between hydrogen-shell and helium-core burning stars 
\citep{Bedding_2011,Stello_2013,Mosser_2014}
and to measure radial differential rotation \citep{Beck_2011,Mosser_2012}. 
This usefulness arises because 
each acoustic non-radial mode in the envelope couples to multiple gravity
modes in the core, forming several observable mixed modes
with frequencies in the vicinity of the acoustic mode \citep{Beck_2011}.
This coupling is strongest for dipole modes, making them the most useful
probes of the core \citep{Dupret_2009}.

\begin{figure*}
\centering
\includegraphics[width=16cm]{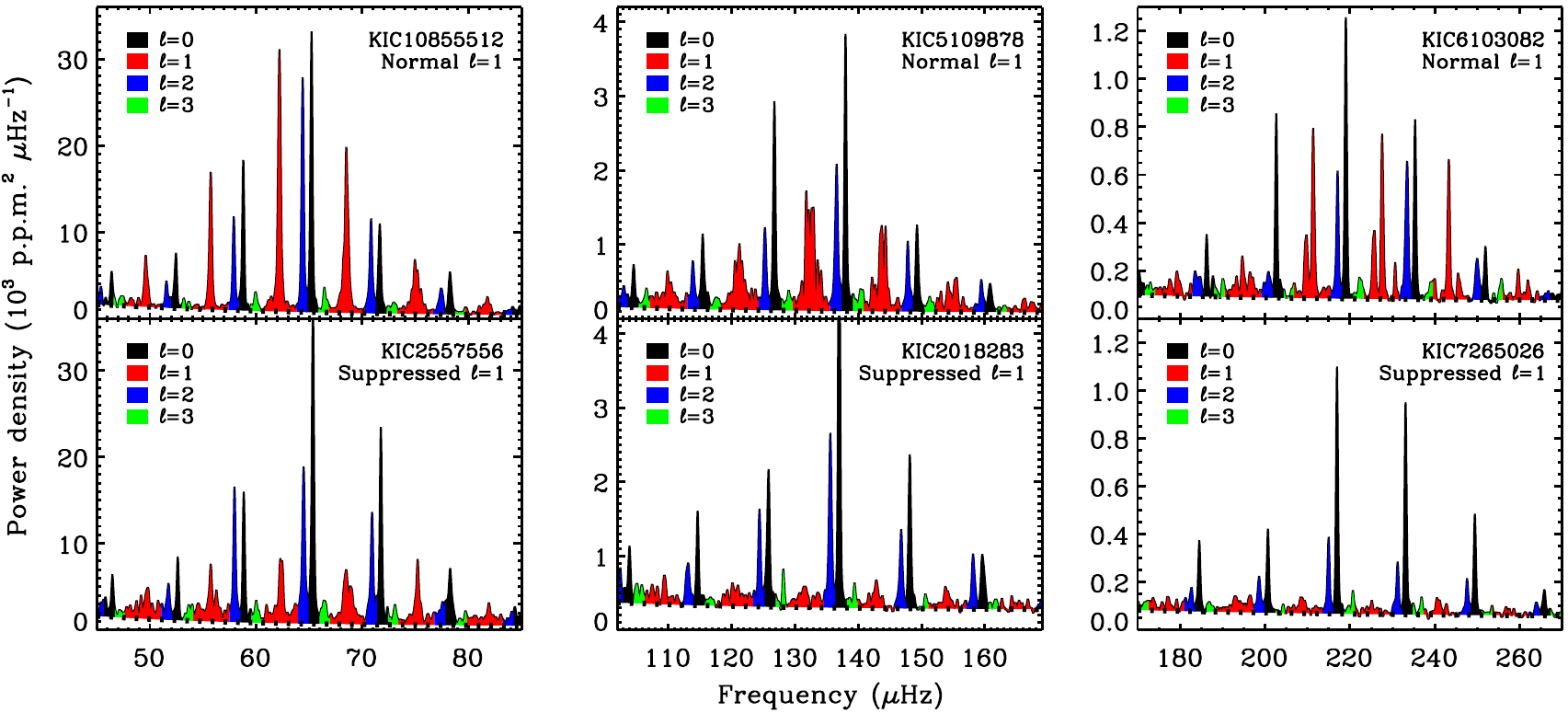}

\caption{Oscillation spectra of six red giants observed with \kepler. The stars are ordered in three pairs, each representing a different evolution stage ranging from the most evolved (lowest oscillation frequencies) on the left to the least evolved (highest frequencies) to the right. The coloured regions mark the power dominated by modes of different degree $\ell=0$--3. For clarity the spectra are smoothed by 3\% of the frequency separation between overtone modes, which for the most evolved stars tend to create a single peak at each acoustic resonance, even if it comprises multiple closely-spaced mixed modes (red peaks in the left and centre panels). The slightly downward sloping horizontal dashed line indicates the noise level. Observations of each star were made during the first 37 months of the \kepler\ mission (observing quarters Q0--Q14).%
}
\end{figure*}

Figure 1 shows the oscillation power spectra of red giants at three
different evolutionary stages observed by NASA's \kepler\ mission. 
For ``normal'' stars (upper panels), the dipolar modes (red peaks) have similar power
to the radial modes (black peaks).  However, at each stage 
of evolution we also find stars with greatly suppressed dipolar modes (lower panels).
Suppressed dipolar modes have been reported in a few dozen red giant stars, with an occurrence rate of about 20\% 
\citep{Mosser_2011,Garc_a_2014}. The cause of this phenomenon has been puzzling until recent theoretical work, which showed that the suppression can be explained if waves entering the stellar core are prevented from returning to the envelope.
This occurs for dipolar modes if there are strong magnetic fields in the
core, giving rise to a ``magnetic greenhouse effect'' 
\citep{Fuller15}.

We measured the amount of suppression by comparing the integrated power of the dipolar and radial modes (the dipole 
mode visibility, $V^2$), averaged over the four orders centred on \numax. 
While the normal stars show dipole mode visibilities of $V^2 \approx 1.5$, independent of \numax,
\citep{Ballot_2011,Mosser_2011},
the stars with suppressed modes have $V^2\approx 0.5$ for \numax\ $\simeq 70\,$\muhz\ and 
down to almost zero for the least evolved red giants oscillating above $200\,$\muhz\ (Fig. 1).  

In Fig. 2 we show the 
dipole mode visibility for about 3600 red giants observed over the first 
37 months of the \kepler\ mission. Our analysis is restricted to a sample of stars with \numax\ 
larger than 50\muhz\ and masses below 2.1\msol\ which, assuming no observational uncertainties, is expected to include only red giants that have not started burning helium in their cores \citep{Stello_2013}.  
We cross-matched our sample with those of known helium-burning stars \citep{Stello_2013,Mosser_2014}, 
which allowed us to identify and remove a small fraction of evolved stars burning helium that, due to measurement uncertainty, had entered our sample ($2\%$ of our sample, almost all with \numax\ $< 70$\muhz).

The stars in Fig. 2 form two distinct branches that gradually merge as the stars
evolve leftwards towards lower \numax. 
Most stars fall on the ``normal'' upper branch of $V^2\approx 1.5$, in
agreement with previous results \citep{Mosser_2011}.    
The lower branch, with suppressed dipole modes, agrees remarkably well with
theoretical predictions (black curve). This prediction
assumes that all the wave energy leaking into the stellar core is trapped by 
a magnetic greenhouse effect caused by strong internal magnetic fields \citep{Fuller15}.
The decrease of the suppression towards lower \numax\ is a consequence of the weaker coupling between acoustic waves in the envelope and gravity waves in the core \citep{Fuller15}.  
With this large sample we have been able to separate the stars in Fig.2 into five different mass intervals, from 0.9 to 2.1\msol. It is striking how 
strongly the relative
population on the lower branch (stars with suppressed dipole modes) depends
on mass.

We quantify the mass dependence in Fig. 3 by
showing the relative number of dipole-suppressed stars (those below the dashed
line in Fig. 2) in narrow mass intervals.  
We see no suppression in red giants below 1.1\msol, which 
coincides with the mass below
which they did not have convective cores during the core-hydrogen-burning phase \citep{Kippenhahn_1990}.
The onset of magnetic suppression above this threshold suggests that at
least some of those stars had
convectively driven magnetic dynamos in their cores during the core-hydrogen-burning (main-sequence)
phase. This is supported by 3D hydrodynamical modeling of these stars
\citep{Brun_2005}. Red giants no longer contain convective cores, leading
us to conclude that the strong magnetic fields in suppressed oscillators are the remnants of the fields produced by core dynamos during the main sequence.

Figure 3 shows that the incidence of magnetic suppression increases with
mass, with red giants above 1.6\msol\ showing a remarkable suppression rate of 50-60\%.
These have evolved from main-sequence A-type stars, among which only up to $\approx 10$\% are observed to have strong fields at their surfaces
\citep{2008CoSka..38..443P}. 
We conclude that these magnetic A stars represent only the tip of the iceberg, and that a much larger fraction of A
stars have strong magnetic fields hidden in their cores.

\begin{figure}
\centering
\includegraphics[width=8.8cm]{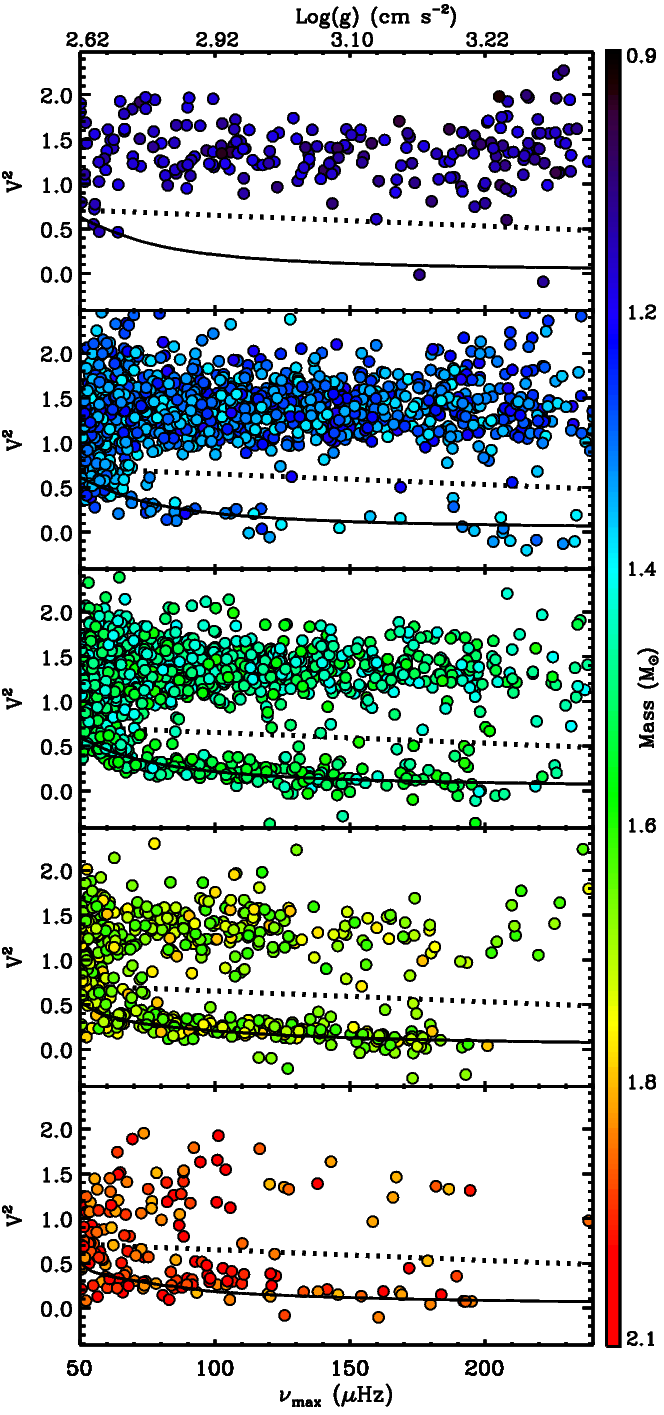}

\caption{Visibility of dipolar modes for red giants observed with \kepler. The abscissa is the central frequency of the oscillations, which correlate closely with surface gravity shown at the top axis. Stars evolve from right to left in the diagram, corresponding roughly to the beginning of the red giant phase to the red giant luminosity bump \citep{Salaris_2005}. The upper limit on \numax\ is set by the sampling of the \kepler\ data. Each panel shows stars in a different mass bracket increasing from top to bottom (indicated by the colour bar annotation on the right). Mass is calculated from asteroseismic scaling relations \citep{Stello_2013}, and has a formal uncertainty of 10\% \citep{Miglio_2011}. The solid black line shows the theoretical predicted visibility of suppressed dipole modes \citep{Fuller15} assuming a stellar mass of 1.6\msol\ and a mode lifetime for radial modes of 20 days \citep{E_Corsaro_2015}. The fiducial dashed line separates the two branches of normal and dipole-suppressed stars.%
}
\end{figure}

\begin{figure}
\centering
\includegraphics[width=8.8cm]{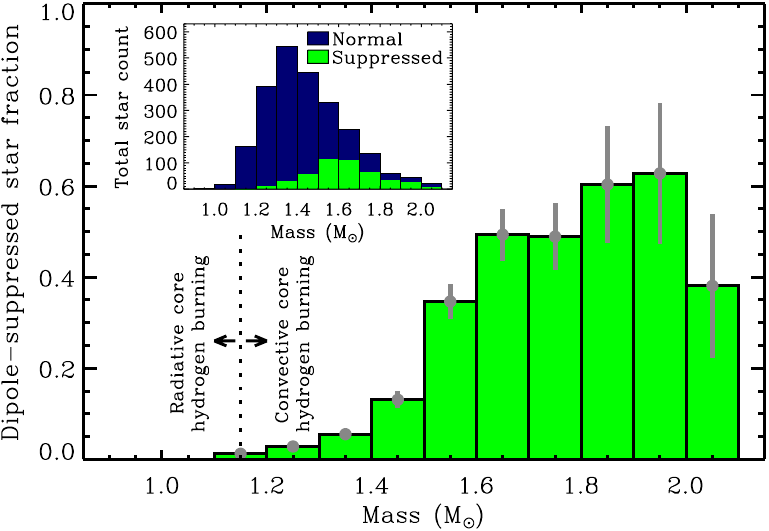}

\caption{Observed fraction of stars with suppressed dipolar modes.  The
  abscissa is the stellar mass (in solar units).  For each mass bin we calculated the dipole-suppressed star
  fraction as the number of stars that fall below the dashed line in
  Fig. 2, relative to all stars in that same mass bin.  
  To make the distinction unambiguously between normal stars and stars with suppressed dipoles, we only counted stars with \numax\ $>70$\muhz.
  The uncertainty in the fractions (grey vertical errorbars) are based on
  Poisson statistics of the total star counts (inset: blue plus green) and of
  the number of dipole-suppressed stars (inset: green).  The vertical dotted line separates stars 
  for which hydrogen-core burning took place in either a radiative or convective environment for solar metallicity
  \citep{Kippenhahn_1990}.%
}
\end{figure}

In Fig. 4 we show the observed \numax\ and inferred
mass of all the stars superimposed on a contour plot of minimum magnetic field strengths required for
mode suppression \citep{Fuller15}. For stars with suppressed modes
(filled red circles), the underlying
colour provides a lower bound to the field strength at the hydrogen-burning
shell. For stars without 
suppressed modes (open black
circles), the underlying colour represents an upper limit to the field at the hydrogen-burning shell;
above or below the shell the field could potentially be larger. Hence, normal and dipole-suppressed stars
that fall in the same regions of Fig. 4 may have core field strengths that are only slightly different.
However, we expect that the dipole-suppressed stars on average exhibit stronger core fields than their
normal counter parts.

Considering again the low-mass stars ($< 1.1$\msol), of which none show
suppression, we see from Figure 4 that magnetic fields above $\approx 10 \,
{\rm kG}$ are not present at the hydrogen-burning shell when the stars are
just below the red giant luminosity bump (\numax\ $\approx 70-100 \, \mu{\rm Hz}$).
Assuming magnetic flux conservation from the main-sequence phase, this
suggests that fields above $\approx 5 \, {\rm kG}$ do not exist within the
cores of Sun-like stars \citep{Fuller15}. 
Large scale fields in the solar interiors have been
discussed in order to explain the properties of the
tachocline \citep{GoughMcIntyre98}. However, our results do
not rule out strong
horizontal fields near the radiative-convective boundary because those
fields would be outside the core and could not cause mode suppression when the star evolves into a red giant.

Turning to higher masses we see that, for a given \numax, stars above
1.4\msol\ require increasingly strong magnetic fields to suppress their
dipolar modes. From Figure 4, there is no clear upper limit to the field
strengths present in red giant cores, given that dipole-suppressed stars are
common even when field strengths $ B> 1 \, {\rm MG}$ are required for
suppression.  However, the hint of a decline in the occurrence of
dipole-suppressed stars above 2\msol\ seen in Fig. 3 suggests there may be
a mass above which dynamo-generated magnetic fields can no longer cause
oscillation mode suppression in intermediate-mass stars.

The high occurrence rate of dipole mode suppression demonstrates that
core-dynamo-generated fields can remain through the red giant phase, more
than $10^8 \, {\rm yr}$ after the dynamo has shut off at the end of
core-hydrogen-burning. This indicates that dynamo-generated fields are frequently
able to settle into long-lived stable configurations, a result that was not
certain from magnetohydrodynamical simulations
\citep{Braithwaite_2004,Braithwaite_2006,Duez_2010}. The occurrence rate of
suppressed dipole modes in intermediate-mass red giants is much higher
than the occurrence rate of
strong fields at the surfaces of the main-sequence A stars from which they
evolved. The latter fields are thought to have
been generated by a pre-hydrogen-core burning dynamo during star formation
\citep{Moss_2004}. We conclude that fields generated during core-hydrogen-burning are able to settle into stable equilibrium configurations much more
commonly (more than $60\%$ of the time) than fields generated during
star formation (less than $10\%$ of the time).

Our results show that main-sequence stars with no observable magnetic field
at the surface can still harbour strong fields in the core that survive
into the red giant phase. 
The presence of internal magnetic fields might play an important role for angular momentum transport.  
Fields too weak to suppress dipolar oscillation modes may exist in normal
red giants, and these fields may nevertheless transport enough angular momentum to help explain the measured rotation rates of red giant cores \citep{Mosser_2012,Cantiello_2014}. 
After some time, intermediate-mass red giants also start burning helium in
their cores. Suppressed dipolar modes in those so-called red clump stars
will reveal whether the fields survive until helium-core burning, and
whether they can account for magnetic fields observed in stellar remnants
such as white dwarfs.  Like intermediate-mass stars, more massive stars
($M>10$\msun) also undergo convective hydrogen-core burning that generates
a magnetic dynamo, and which may produce the magnetic fields observed in
many neutron stars.

\begin{figure}
\centering
\includegraphics[width=9.0cm]{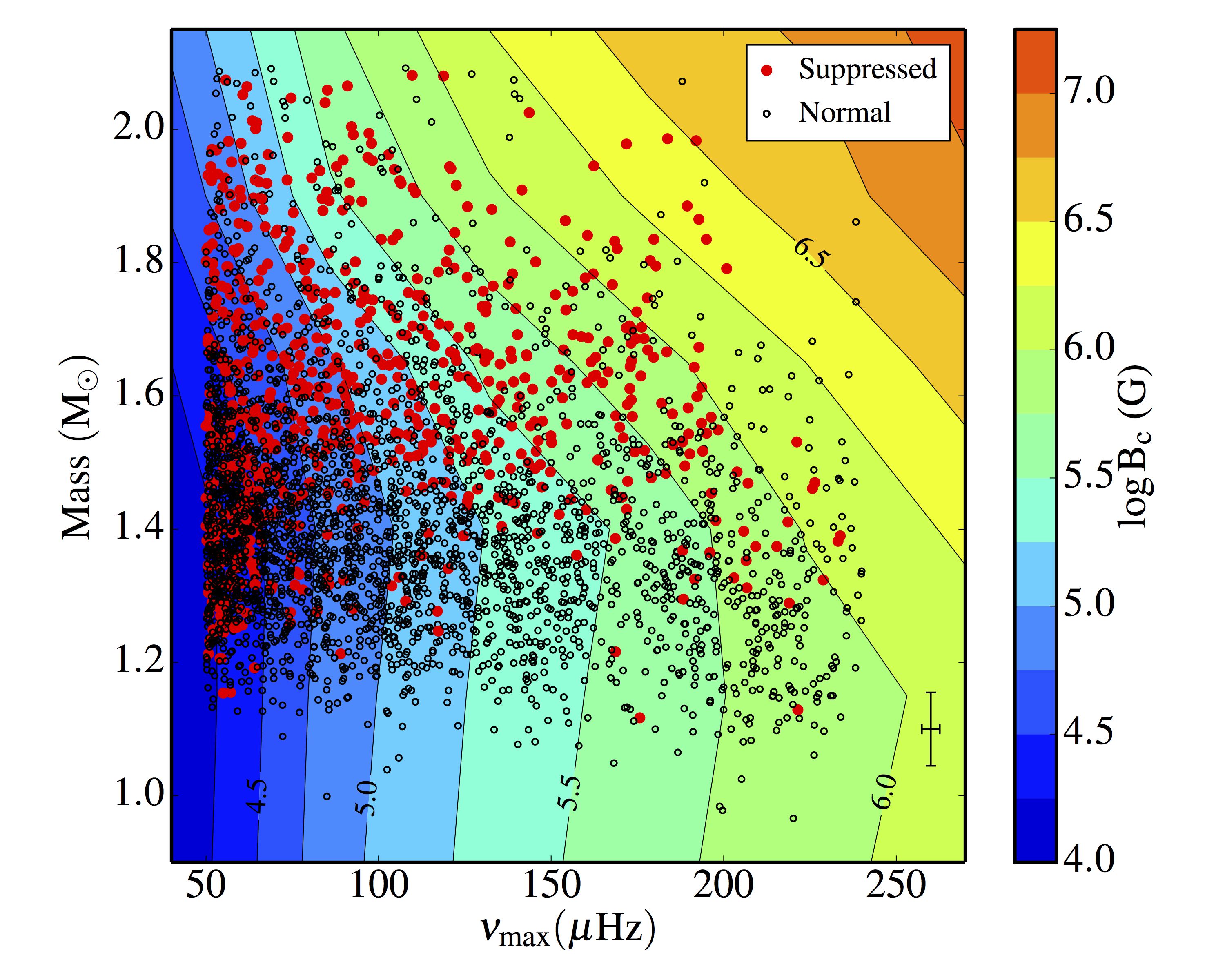}

\caption{Critical magnetic field strength required to suppress dipole mode oscillations. The abscissa is the
    observed central frequency of the oscillations.  The ordinate is the
    inferred asteroseismic mass.  The coloured contours indicate the minimum
    magnetic field at the hydrogen shell required for
    mode suppression (the critical field, $B_c$).  Filled red
    circles mark  stars with observed suppressed modes, and open circles
    mark normal (not suppressed) stars. The cross shows a typical errorbar for the data points.   
    The uncertainty in $B_c$ due to uncertainty in mass is negligible for stars below 1.4\msol\ and is no
    more than 25\% for the more massive stars. %
}
\end{figure}

\bigskip

\small
{\sffamily

%\bibliography{converted_to_latex_final}

\section*{Acknowledgments} 

This paper has been written collaboratively, on the web, using Authorea.
We acknowledge the entire {\it Kepler} team, whose efforts made
these results possible.
D.S. is the recipient of an Australian Research Council Future Fellowship
(project number FT140100147). 
J.F. acknowledges support from NSF under grant no. AST-1205732 and
through a Lee DuBridge Fellowship at Caltech. R.A.G. acknowledge the support
of the European Community{'}s Seventh Framework Programme (FP7/2007-2013)
under grant agreement No. 269194 (IRSES/ASK), and from the CNES. 
D.H. acknowledges support by the Australian Research Council's Discovery
Projects funding scheme (project number DE140101364) and support by the
National Aeronautics and Space Administration under Grant NNX14AB92G 
issued through the Kepler Participating Scientist Program.
This project was supported by NASA under TCAN grant number NNX14AB53G, and
the NSF under grants PHY 11-25915 and AST 11-09174.  
Funding for the Stellar Astrophysics Centre is provided by The Danish
National Research Foundation (Grant agreement no.: DNRF106). The research
is supported by the ASTERISK project (ASTERoseismic Investigations with
SONG and Kepler) funded by the European Research Council (Grant agreement
no.: 267864). 

\section*{Author contributions} 
D.S. measured and interpreted mode visibilities; M.C. and J.F. calculated
and interpreted theoretical models;  D.H. and D.S. calculated power spectra
and measured large frequency separations; R.A.C., T.R.B., L.B., and
V.S.A. contributed to the discussion of the results. All authors commented
on the manuscript. 

\section*{Author information} 
Reprints and permissions information is available at www.nature.com/reprints
The authors declare no competing financial interests. Correspondence and
requests for materials should be addressed to D.S. (stello@physics.usyd.edu.au).

\end{document}